\documentclass[intlimits,twoside,a4paper]{article}
\usepackage{gensymb}
\usepackage{graphicx}

\usepackage[cp1251]{inputenc}

\usepackage[eqsecnum]{cmpj3}

\issue{2017}{20}{2}{23705}
\doinumber{10.5488/CMP.20.23705}
\title[Investigation on $\alpha$-MoO$_3$ nanostructures]%
{Interaction of NH$_3$ gas on $\alpha$-MoO$_3$ nanostructures --- a DFT investigation }%
\author[V. Nagarajan, R. Chandiramouli]{V. Nagarajan, R. Chandiramouli\footnote{Corresponding author}}
\address{School of Electrical and Electronics Engineering, Shanmugha Arts Science Technology and Research Academy (SASTRA) University, Tirumalaisamudram, Thanjavur, Tamil nadu --- 613 401, India
}

\date{Received February 21, 2017, in final form April 18, 2017}
\begin{document}

\maketitle

\begin{abstract}
The structural stability, electronic properties and NH$_3$ adsorption properties of pristine, Ti, Zr and F substituted $\alpha$-MoO$_3$ nanostructures are successfully studied using density functional theory with B3LYP/LanL2DZ basis set. The structural stability of $\alpha$-MoO$_3$ nanostructures is discussed in terms of formation energy. The electronic properties of pristine, Ti, Zr and F incorporated $\alpha$-MoO$_3$ nanostructures are discussed in terms of HOMO-LUMO gap, ionization potential and electron affinity. $\alpha$-MoO$_3$ nanostructures can be fine-tuned with suitable substitution impurity to improve the adsorption characteristics of ammonia, which can be used to detect NH$_3$ in a mixed environment. The present work gives an insight into tailoring $\alpha$-MoO$_3$ nanostructures for NH$_3$ detection.
\pacs 71.15.Mb
\end{abstract}

\section{Introduction}
Ammonia (NH$_3$) is widely used in automobiles, food industry and agriculture in the form of fuel, antimicrobial agent and fertilizers \cite{1}. The detection of NH$_3$ is a significant criterion due to its hazard. The exposure limit of ammonia is around 25~ppm, which is recommended by Occupational Safety and Health Administration (OSHA) \cite{2}. There are several analytical techniques adopted to detect ammonia gas, which include the laser methods \cite{3}, electrochemical methods, optical methods \cite{4} and mass spectrometry~\cite{5}. These techniques are time-consuming and also need sophisticated instruments. In this context, an inexpensive and real-time sensor is required to detect trace the amounts of ammonia at ppm level.
Metal oxide semiconductor (MOS) thin films are extensively used for gas sensors as their resistivity changes upon interaction with toxic gas molecules \cite{6,7}. Moreover, MOS sensors are easy to fabricate, low cost and are uniform in performance among other types of gas sensors. Besides, the morphology such as rods, belts and wires in the micro-dimension and nano-dimension show a significant performance in MOS sensors. Furthermore, these types of nanostructured materials offer high surface-to-volume ratio, which leads to an enhanced performance in gas sensing \cite{8,9,10}.

	Among various metal oxide semiconductors, molybdenum oxide (MoO$_3$) is an excellent candidate for electrochromic, catalytic and gas sensing applications. MoO$_3$ is n-type semiconductor with a wide band gap; the conductivity arises due to oxygen vacancies. The band gap of MoO$_3$ is found to be around 2.69--2.76~eV \cite{11}. MoO$_3$ is also used as catalyst for the reduction of NO$_x$ in the petroleum and chemical industry and oxidation of hydrocarbons \cite{12,13,14,15}. Moreover, there are reports for enhancing the gas sensing properties of molybdenum oxide based device to detect LPG \cite{16}, CO \cite{17,18}, NH$_3$ \cite{13,16,19}, H$_2$ \cite{16,17}. Furthermore, the synthesis of MoO$_3$ includes thermal evaporation \cite{20}, pulsed laser deposition \cite{13}, sol-gel \cite{1,21}, electro-deposition \cite{22} and chemical vapor deposition \cite{23,24}.

	MoO$_3$ exhibits in three polymorphic phases \cite{25} namely stable orthorhombic $\alpha$-MoO$_3$,  meta-stable monoclinic $\beta$-MoO$_3$, hexagonal $\beta$-MoO$_3$.  Kannan et al. \cite{26} have reported the influence of the precursor solution volume on the properties of spray deposited $\alpha$-MoO$_3$ thin films. Martinez et al. \cite{27} have studied the gas sensing properties of spray deposited MoO$_3$ thin films. Hussain et al. \cite{28} have reported activated reactive evaporated MoO$_3$ thin films for gas sensor applications and they observed that $\alpha$-MoO$_3$ are capable of sensing both CO and NH$_3$ gases at below 10~ppm of concentration in dry air. Density functional theory (DFT) is an efficient method for studying the interaction between compounds and adsorption characteristics of compounds \cite{29,30,31}. Based on these aspects, literature survey was conducted using CrossRef metadata search and it is inferred that not much work was reported based on DFT methods to investigate the adsorption properties of NH$_3$ on $\alpha$-MoO$_3$ nanostructures. The motivation behind the present work is to improve the NH$_3$ adsorption properties on $\alpha$-MoO$_3$ nanostructures with the incorporation of dopants. The novel aspect of this work is to study the adsorption characteristics of NH$_3$ on $\alpha$-MoO$_3$ nanostructures with the substitution of Ti, Zr and F as substitution impurities.

\section{Computational methods}
In the present work, Gaussian 09 package \cite{32} is used to optimize the pristine, Ti, Zr and F substituted $\alpha$-MoO$_3$ nanostructures. This package is also used to investigate the adsorption properties of NH$_3$ gas molecules on $\alpha$-MoO$_3$ base material. The calculations were carried out for isolated MoO$_3$ base material with periodic boundary condition (PBC). Moreover, the atoms were fixed along the direction perpendicular to the molecular plane and allowed us to relax along the other planes. In the present work, DFT is utilized in accordance with Becke’s three-parameter hybrid functional in combination with Lee-Yang-Parr correlation functional (B3LYP)/LanL2DZ basis set \cite{33,34,35,36}. The selection of a suitable basis set is an important criterion for optimizing $\alpha$-MoO$_3$ nanostructures. In our previous study we demonstrated a density functional theory with the use of all-electron basis sets, but methods including effective core potentials (ECPs) are good in reducing the computational cost. Furthermore, the efforts have been taken into account for measuring the performance of basis set in order to approximate the same set of density functional. Moreover, we used ECP basis set such as LanL2DZ (Los Alamos National Laboratory 2 Double-Zeta), which is more suitable for transition metals. Besides, utilization of all-electron basis sets for the remaining non-transition metal elements has become more popular in computational studies on transition metal containing materials. Further, the atomic number of molybdenum and oxygen is forty two and eight, respectively. In the present work,  $\alpha$-MoO$_3$ nanostructures are studied with the incorporation of impurities such as Ti, Zr and F. It is known that Ti and Zr belongs to group IVB and is 4th and 5th period, respectively. Thus, LanL2DZ effective core potential will be suitable for the optimization of $\alpha$-MoO$_3$ nanostructures with impurities. In addition, LanL2DZ basis set is applicable for the elements such as H, Li-La and Hf-Bi, which gives the best results with the pseudopotential approximation. Hence, LanL2DZ basis set is a suitable basis set to optimize $\alpha$-MoO$_3$ with pseudopotential approximation \cite{37,38,39}. The HOMO-LUMO gap and density of states spectrum (DOS) of $\alpha$-MoO$_3$ nanostructures are calculated using Gauss Sum~3.0 package \cite{40}. The energy convergence is obtained within the range of 10--5~eV, during the optimization of $\alpha$-MoO$_3$ nanostructures.

\section{Results and discussion}
The present work mainly focuses on the study of ionization potential (IP), HOMO-LUMO gap, dipole moment, electron affinity (EA), Mulliken population and adsorption properties of NH$_3$ gas molecules in MoO$_3$ base material with the incorporation of impurities such as Ti, Zr and F in $\alpha$-MoO$_3$ nanostructures. The reason behind the selection of Ti, Zr and F as impurities is that Ti and Zr belong to the transition metals like Mo. The electronic and structural properties of $\alpha$-MoO$_3$ nanostructure can be controlled and fine-tuned with chemical modifications, such as doping \cite{41}. Especially, the doping methodology has been widely utilized for numerous conjugated polymers, while dopants play a vital role in modifying their electronic properties of conventional covalent semiconductors \cite{42}. This doping technique can be performed either by application of an external electric field or charge transfer. Owing to the application of doping mechanism, optical and electronic properties of the respective conducting polymers can be widely engineered. In addition, the doping effect influences the transfer charge from semiconductor to metal or insulator based on the concentration of a dopant \cite{43}.
The electronic properties of conjugated polymers can be easily fine-tuned  either by atomic/molecular scale doping or chemical modification \cite{41}. Kaloni et al. \cite{44,45} have reported the effect of the doping mechanism on polythiophene and polypyrrole based materials and studied the structural and electronic properties of pure and doped polymers. In the present work, $\alpha$-MoO$_3$ nanostructures get distorted, while optimizing the base material with Ti, Zr and F elements as dopant, which results in the variation of electronic properties. Moreover, the conductivity of $\alpha$-MoO$_3$ base material changes with the substitution of impurities. In addition, fluorine is abundant in electrons compared to oxygen, which leads to an increase in n-type behavior. Based on these aspects, the dopants are selected to incorporate in $\alpha$-MoO$_3$ base material and these impurities tune the conducting property in $\alpha$-MoO$_3$ base material. Therefore, the adsorption properties of NH$_3$ on $\alpha$-MoO$_3$ nanostructures can be improved with the substitution impurities.

\begin{figure}[!t]
\begin{center}
\includegraphics[width=0.8\textwidth]{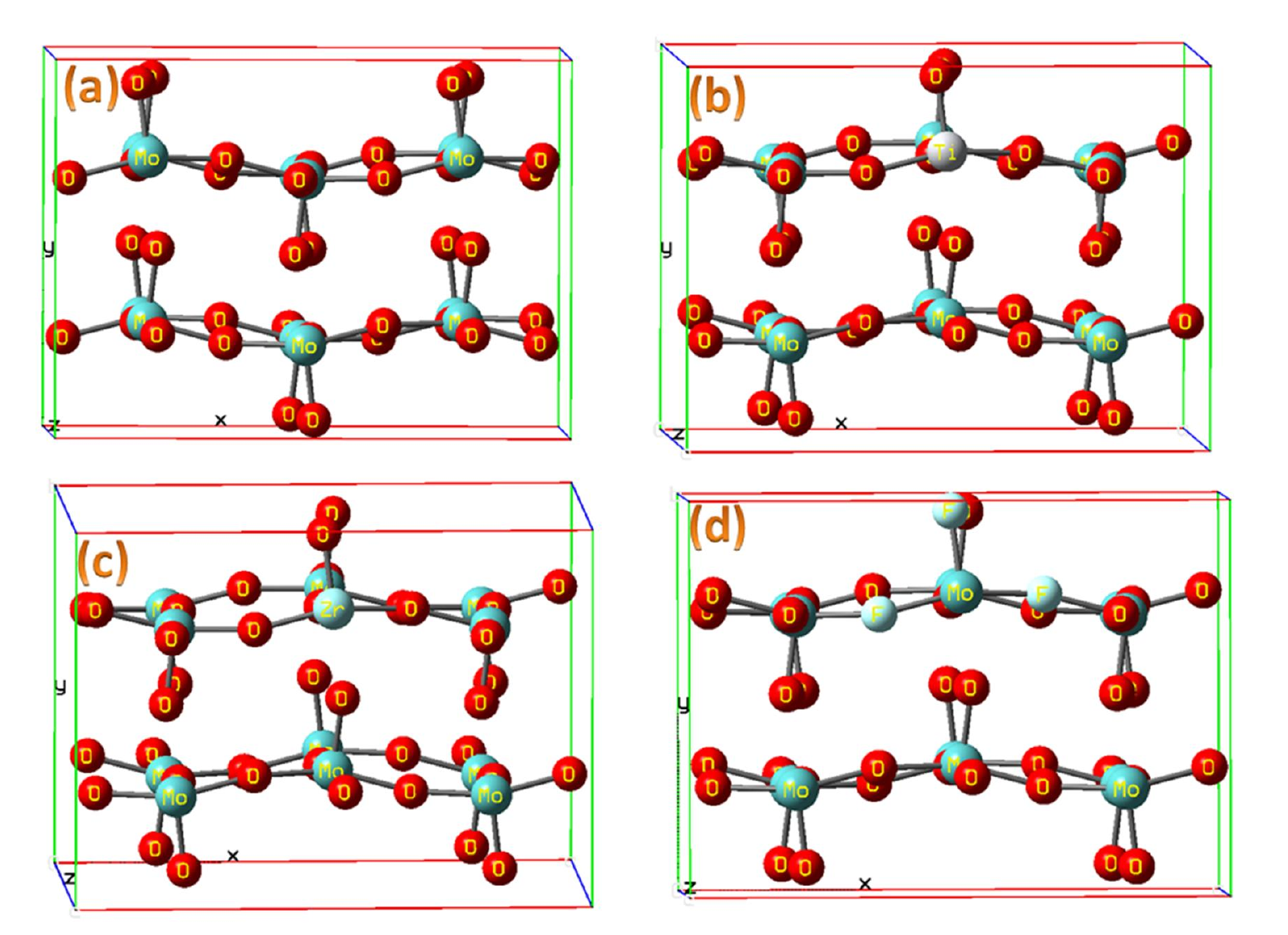}
\caption{(Color online) (a) Pristine, (b) Ti substituted, (c) Zr substituted and (d) F substituted $\alpha$-MoO$_3$ nanostructure.} \label{fig1}
\end{center}
\end{figure}

Figure~\ref{fig1} represents the structure of pristine, Ti, Zr and F substituted $\alpha$-MoO$_3$ nanostructures, respectively. The structure of $\alpha$-MoO$_3$ is taken from International Centre for Diffraction Data (ICDD) card number: 89-7112. The pristine $\alpha$-MoO$_3$ nanostructure has twelve Mo atoms and thirty eight O atoms. Ti incorporated $\alpha$-MoO$_3$ nanostructure has eleven Mo atoms, thirty eight O atoms and one Mo atom is replaced with one Ti atom. Similarly, Zr substituted $\alpha$-MoO$_3$ nanostructure has eleven Mo atoms, thirty eight O atoms and one Mo atom is replaced with one Zr atom. In the case of F substituted $\alpha$-MoO$_3$ nanostructure, it has twelve Mo atoms, thirty five O atoms and three O atoms are replaced with three F atoms for enhancing the adsorption properties of NH$_3$ on $\alpha$-MoO$_3$ base material.

\subsection{Structural stability and electronic properties of $\alpha$-MoO$_3$ nanostructures}

The structural stability of pristine, Ti, Zr and F substituted $\alpha$-MoO$_3$ nanostructures can be described in terms of formation energy as shown in equation~(\ref{3.1})
\begin{equation}
\label{3.1}
	E_{\text{form}} = 1/n[E (\alpha\text{-MoO}_3 \,\,\text{nanostructure})- pE(\text{Mo}) - qE(\text{O}) - rE(\text{dopant})],
\end{equation}
where $E(\alpha$-MoO$_3$ nanostructures) refers to the total energy of $\alpha$-MoO$_3$ nanostructures, $E$(Mo), $E$(O) and $E$(dopant) represent the corresponding energy of isolated Mo, O and dopant atoms namely Ti, Zr and F. $p$, $q$ and $r$ represent the total number of Mo, O and dopant atoms, respectively, and $n$ is the total number of atoms in $\alpha$-MoO$_3$ nanostructure. The dipole moment, point symmetry and the formation energy of pristine, Ti, Zr and F substituted $\alpha$-MoO$_3$ nanostructures are tabulated in table~\ref{table 1}. The formation energy of pristine, Ti, Zr and F substituted $\alpha$-MoO$_3$ nanostructures is $-4.38$, $-4.36$, $-4.30$ and $-4.40$~eV, respectively.

\begin{table}[!t]
  \renewcommand{\arraystretch} {1.3}
  \caption{ Formation energy, dipole moment and point group of $\alpha$-MoO$_3$ nanostructures.}
  \vspace{2ex}
  \label{table 1}
  \centering
  \begin{tabular}{|c|c|c|c|}
  \hline\hline
  Nanostructures	&Formation energy &	Dipole moment & Point group \\
     & (eV)&(D)	 & \\
  \hline\hline
 Pristine  $\alpha$-MoO$_3$  nanostructure	&$-4.38$	&5.62	&$C_1$ \\
\hline
  Ti substituted  $\alpha$-MoO$_3$  nanostructure	&$-4.36$	&19.18	&$C_1$ \\
  \hline
	Zr substituted  $\alpha$-MoO$_3$   nanostructure	&$-4.30$&	35.08	&$C_1$\\
	\hline
	F substituted  $\alpha$-MoO$_3$  nanostructure	&$-4.40$&	40.47	&$C_1$ \\
	\hline\hline
  \end {tabular}
  \end{table}
	
Before studying the adsorption characteristics, the structural stability of $\alpha$-MoO$_3$ base material must be studied. The structural stability of $\alpha$-MoO$_3$ nanostructures slightly decreases with the substitution of Ti and Zr. The formation energy of Ti and Zr substituted $\alpha$-MoO$_3$ nanostructures is relatively low compared to pristine $\alpha$-MoO$_3$ nanostructures. By contrast, the stability of $\alpha$-MoO$_3$ nanostructure slightly increases with the substitution of F. The dipole moment (DP) gives a clear picture about the distribution of charges in $\alpha$-MoO$_3$ nanostructure. The corresponding dipole moment value of pristine, Ti, Zr and F incorporated $\alpha$-MoO$_3$ nanostructures is 5.62, 19.18, 35.08 and 40.47~D. Low value of DP is recorded for pristine $\alpha$-MoO$_3$ nanostructure. It infers that the charge distribution is almost uniform, but not in the case of impurity substituted $\alpha$-MoO$_3$ nanostructures. Furthermore, $C_1$ point groups are observed for all $\alpha$-MoO$_3$ nanostructures, which only exhibit identical operation.

The electronic properties of pristine, Ti, Zr and F incorporated $\alpha$-MoO$_3$ nanostructures can be discussed in terms of the highest occupied molecular orbital (HOMO) and the lowest unoccupied molecular orbital (LUMO) \cite{46,47}. In the present work, the electronic properties and adsorption properties of NH$_3$ on $\alpha$-MoO$_3$ nanostructures are studied for a small cluster. The HOMO-LUMO gap of pristine, Ti, Zr and F substituted $\alpha$-MoO$_3$ nanostructures is 4.68, 2.03, 2.22 and 1.9~eV, respectively. The deviation in the HOMO-LUMO gap between experiment and theoretical values arises owing to the selection of the basis set \cite{11}. Moreover, DFT method is broadly linked to the ground state. Thus, the exchange-correlation potential between the excited electronic states may be underestimated. The HOMO-LUMO gap of $\alpha$-MoO$_3$ nanostructure decreases with the substitution of Ti, Zr and F. Thus, the electronic configuration of Mo, Zr, F and Ti element differs, the occupied states also differ. The corresponding electronic configuration of Ti, Zr and Mo is [$1s2$ $2s2$ $2p6$ $3s2$ $3p6$ $3d2$ $4s2$], [$1s2$ $2s2$ $2p6$ $3s2$ $3p6$ $3d10$ $4s2$ $4p6$ $4d2$ $5s2$] and [$1s2$ $2s2$ $2p6$ $3s2$ $3p6$ $3d10$ $4s2$ $4p6$ $4d5$ $5s1$]. In the case of Ti and Zr atom, the outermost $s$-orbital is completely filled with electrons and the electrons are partially occupied in $d$-orbital. Therefore, the $s$ and $d$-states can be chosen as occupied and unoccupied state, respectively. Moreover, for Zr atom, the $d$-state is unfilled. In the case of Mo atom, the outermost $s$-orbital and $d$-orbital are not completely filled with electrons. Thus, due to the doping effects of Ti, Zr and F, the orbital overlapping leads to the variation in the energy gap in $\alpha$-MoO$_3$ nanostructures. The HOMO-LUMO level and the energy gap of $\alpha$-MoO$_3$ nanostructures are tabulated in table~\ref{table 2}.

\begin{table}[!t]
\renewcommand{\arraystretch} {1.3}
\caption{ Adsorption energy, Mulliken population, HOMO-LUMO gap and average energy gap variation of $\alpha$-MoO$_3$ nanostructures.}
\vspace{2ex}
\label{table 2}
\centering
\begin{tabular}{|c|c|c|c|c|c|c|c|}
\hline\hline
$\alpha$-MoO$_3$ nanostructures	&$E_{\text{ad}}$ 	&$Q$ (e)	&$E_{\text{HOMO}}$	&$E_{\text{FL}}$ (eV)	&$E_{\text{LUMO}}$	&$E_{\text g}$ (eV)	&$E_{\text g}^{\text a}$ (\%) \\
\hline\hline
Pristine $\alpha$-MoO$_3$	&$-$	&$-$	&$-$11.13	&$-$8.79	&$-$6.45	&4.68	&$-$	\\
\hline
A	&1.09	&0.30	&$-$11.04	&$-$8.755	&$-$6.47	&4.57	&2.41 \\
\hline
B	&$-$2.45	&0.11	&$-$11.04	&$-$8.685	&$-$6.33	&4.71	&0.64 \\
\hline
C	&$-$8.16	&0.93	&$-$8.55	&$-$6.95	&$-$5.35	&3.2	&46.25 \\
\hline
D	&$-$11.70	&0.56	&$-$9.13	&$-$8.195	&$-$7.26	&1.87	&150.27 \\
\hline
Ti substituted $\alpha$-MoO$_3$	&$-$	&$-$	&$-$10.13	&$-$9.115	&$-$8.1	&2.03	&$-$ \\
\hline
E	&$-$0.27	&0.19	&$-$9.46	&$-$8.785	&$-$8.11	&1.35	&50.37 \\
\hline
F	&$-$2.72	&0.10	&$-$10.34	&$-$9.275	&$-$8.21	&2.13	&4.69 \\
\hline
Zr substituted $\alpha$-MoO$_3$	&$-$	&$-$	&$-$10.48	&$-$9.37	&$-$8.26	&2.22	&$-$ \\
\hline
G	&$-$1.09	&0.13	&$-$10.15	&$-$9.22	&$-$8.29	&1.86	&19.35 \\
\hline
H	&$-$3.26	&0.08	&$-$10.37	&$-$9.31	&$-$8.25	&2.12	&4.72 \\
\hline	
F substituted $\alpha$-MoO$_3$	&$-$	&$-$	&$-$7.9	&$-$6.95	&$-$6	&1.9	&$-$ \\
\hline
I	&$-$14.42	&1.14	&$-$8.22	&$-$6.865	&$-$5.51	&2.71	&29.89 \\
\hline
J	&$-$9.52	&0.67	&$-$9.12	&$-$7.385	&$-$5.65	&3.47	&45.24 \\
\hline\hline
\end {tabular}
\end{table}
	
\begin{figure}[!t]
\begin{center}
\includegraphics[width=0.8\textwidth]{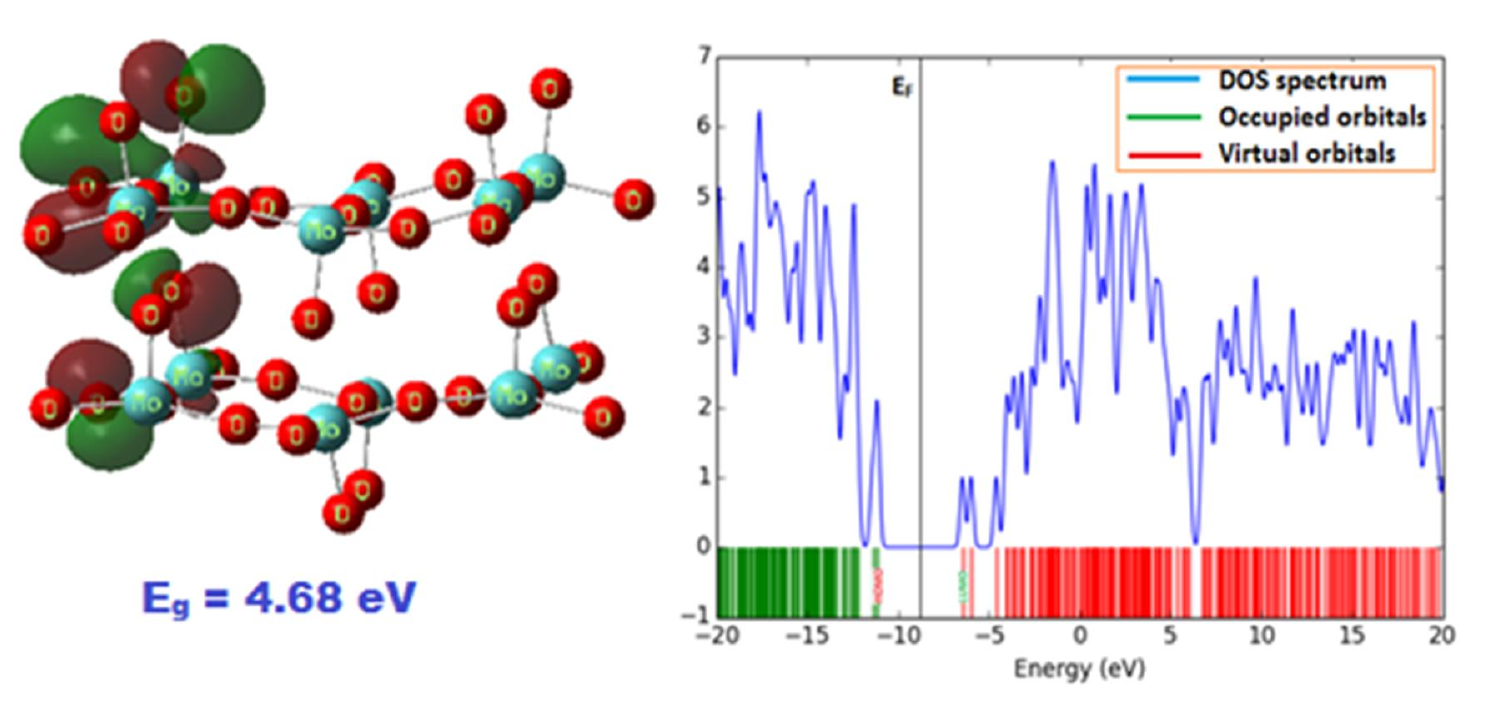}
\vspace{-4mm}
\caption{(Color online) HOMO-LUMO gap and density of states of pristine $\alpha$-MoO$_3$ nanostructure. } \label{fig2}
\end{center}
\end{figure}

\begin{figure}[!t]
\begin{center}
\includegraphics[width=0.8\textwidth]{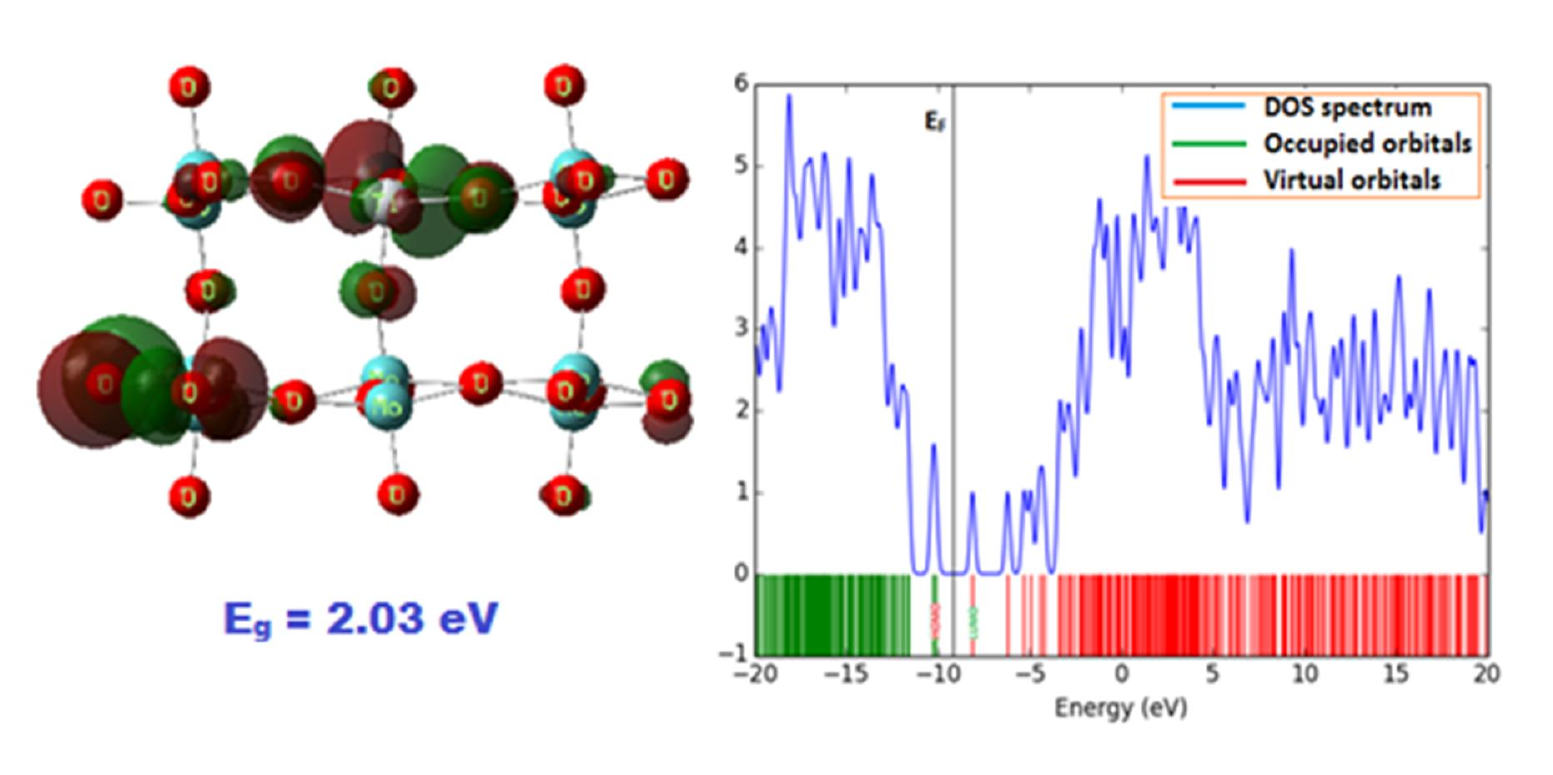}
\caption{(Color online) HOMO-LUMO gap and density of states of Ti substituted $\alpha$-MoO$_3$ nanostructure.} \label{fig3}
\end{center}
\end{figure}

\begin{figure}[!t]
\begin{center}
\includegraphics[width=0.8\textwidth]{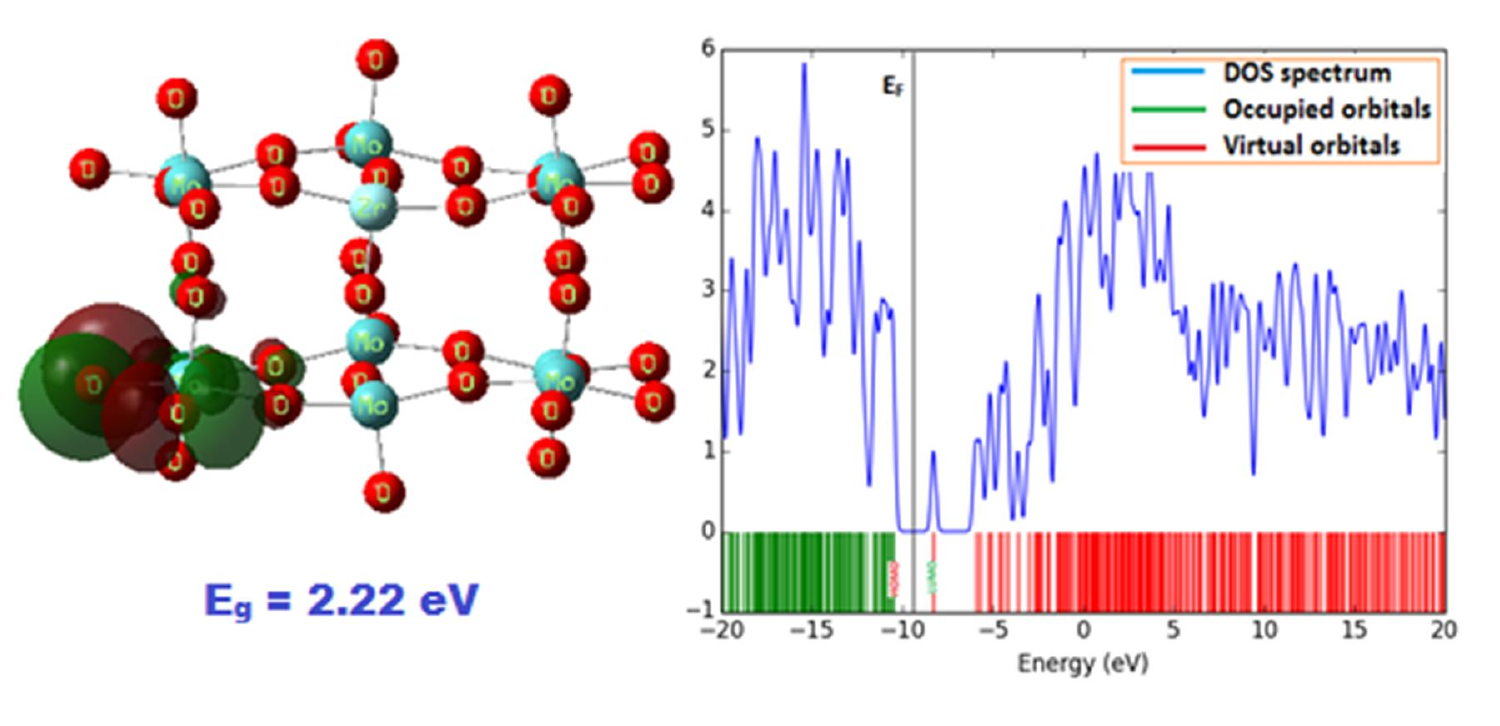}
\caption{(Color online) HOMO-LUMO gap and density of states of Zr substituted $\alpha$-MoO$_3$ nanostructure.} \label{fig4}
\end{center}
\end{figure}

\begin{figure}[!t]
\begin{center}
\includegraphics[width=0.8\textwidth]{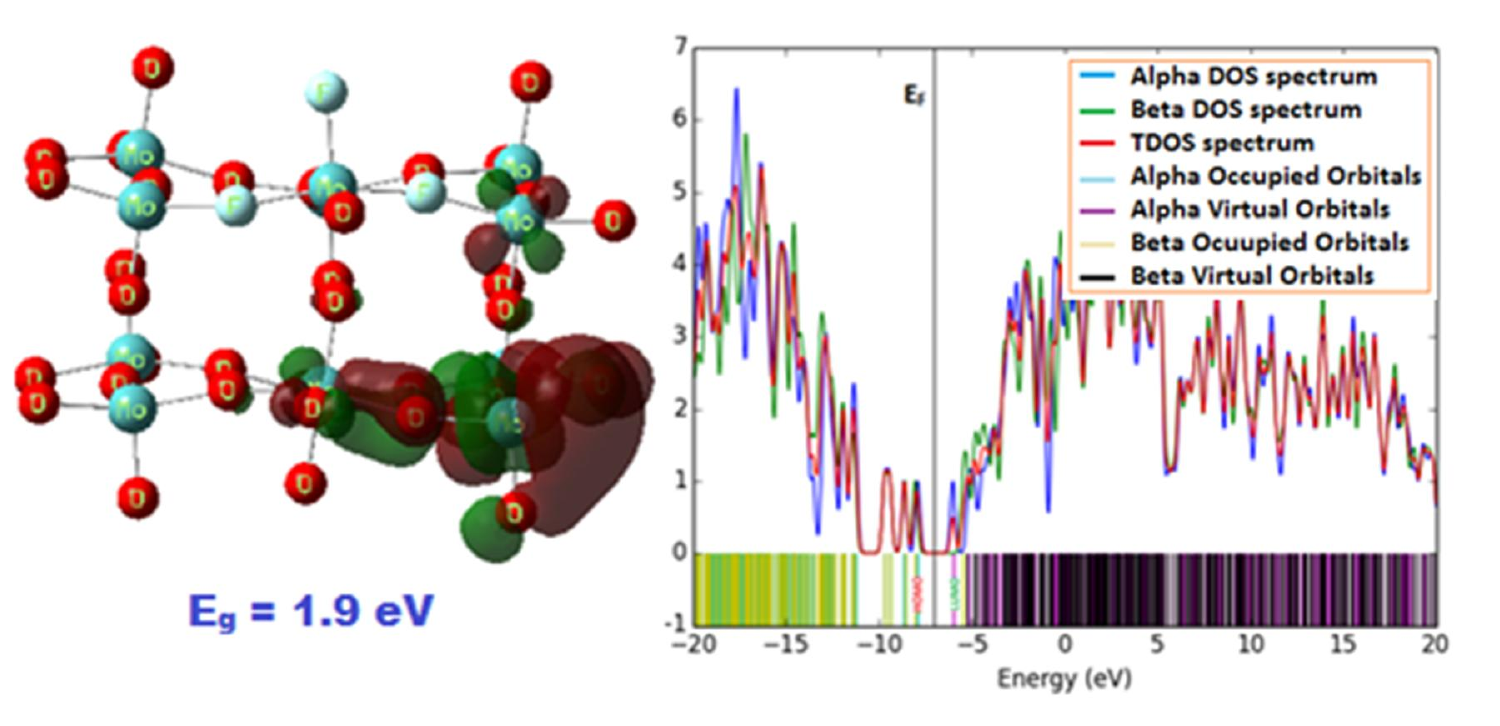}
\caption{(Color online) HOMO-LUMO gap and density of states of F substituted $\alpha$-MoO$_3$ nanostructure.} \label{fig5}
\end{center}
\end{figure}

The density of states spectrum (DOS) gives an insight into the localization of charges in various energy intervals in $\alpha$-MoO$_3$ nanostructures. The DOS spectrum and visualization of HOMO-LUMO gap of pristine, Ti, Zr and F substituted $\alpha$-MoO$_3$ nanostructure are shown in figures~\ref{fig2}--\ref{fig5}, respectively. In the present work, the localization of charges is recorded to be more at LUMO level than at HOMO level, which is confirmed by more peak maxima at LUMO level. More peak maxima in $\alpha$-MoO$_3$ nanostructures arise due to the orbital overlapping of Mo atoms and O atoms in $\alpha$-MoO$_3$ base material. Moreover, the peak maxima in the virtual orbitals of $\alpha$-MoO$_3$ base material are more favorable for adsorption characteristics, since the transfer of electrons between NH$_3$ molecules and virtual orbitals can take place easily.


\begin{figure}[!t]
\begin{center}
\includegraphics[width=0.65\textwidth]{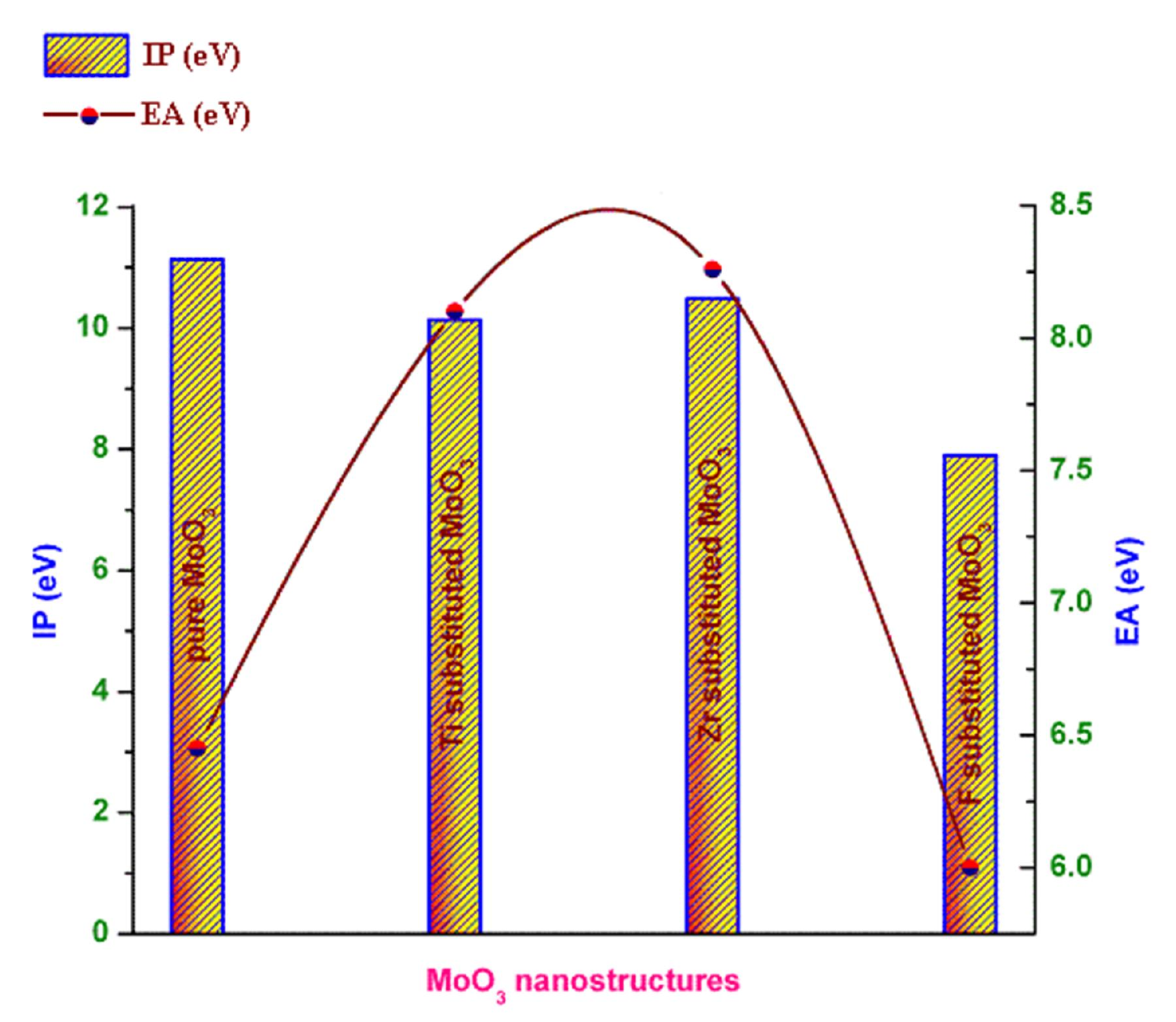}
\caption{(Color online) IP and EA of $\alpha$-MoO$_3$ nanostructure.} \label{fig6}
\end{center}
\end{figure}

The electronic properties of $\alpha$-MoO$_3$ nanostructure can also be discussed with the ionization potential~(IP) and electron affinity~(EA) \cite{48,49}. Figure~\ref{fig6} represents the electron affinity and ionization potential of $\alpha$-MoO$_3$ nanostructures. Generally, IP depicts the amount of energy required to remove the electron from $\alpha$-MoO$_3$ nanostructures and the EA represents the energy change with the addition of electrons in $\alpha$-MoO$_3$ nanostructures. The high value of ionization potential infers that the electrons are tightly bounded to the nucleus in $\alpha$-MoO$_3$ nanostructure. Therefore, pristine, Ti and Zr substituted $\alpha$-MoO$_3$ nanostructures have a high value of ionization potential, which infers that the electrons are strongly attracted to the nucleus and more energy is required to remove electrons from $\alpha$-MoO$_3$ nanostructure. The low value of IP is recorded for F substituted $\alpha$-MoO$_3$ nanostructure. Thus, less energy is required to remove an electron from F substituted $\alpha$-MoO$_3$ nanostructure. Electron affinity plays an important role in plasma physics and in chemical sensors.

The EA value of pristine, Ti, Zr and F incorporated $\alpha$-MoO$_3$ nanostructures is 6.45, 8.1, 8.26 and 6~eV, respectively, which is a favourable condition for chemical sensors. Usually, the amount of transfer of electrons depends on EA of the target gas molecule with the base material. Moreover, EA applies to the electronically conducting solid base material where it can be related to the position of Fermi energy level and the work function. If the base material has a high value of work function,  it will in turn influence EA and it will act as electron acceptor and vice versa. Thus, the base material, which has low value of EA will partially transfer electrons between the target gas molecules and the conduction band of metal oxides. In the chemiresistor type of gas sensors, it is only due to the transfer of electrons that the change in the resistance is observed.  However, in the present work EA for pristine and impurity substituted MoO$_3$ nanostructures, all have EA value of 6 to 8.26~eV. Furthermore, it can be suggested that MoO$_3$ nanostructures can be used as a chemical sensor.

\subsection{Adsorption characteristics of NH$_3$ on $\alpha$-MoO$_3$ nanostructures}

\begin{figure}[!t]
\begin{center}
\includegraphics[width=0.8\textwidth]{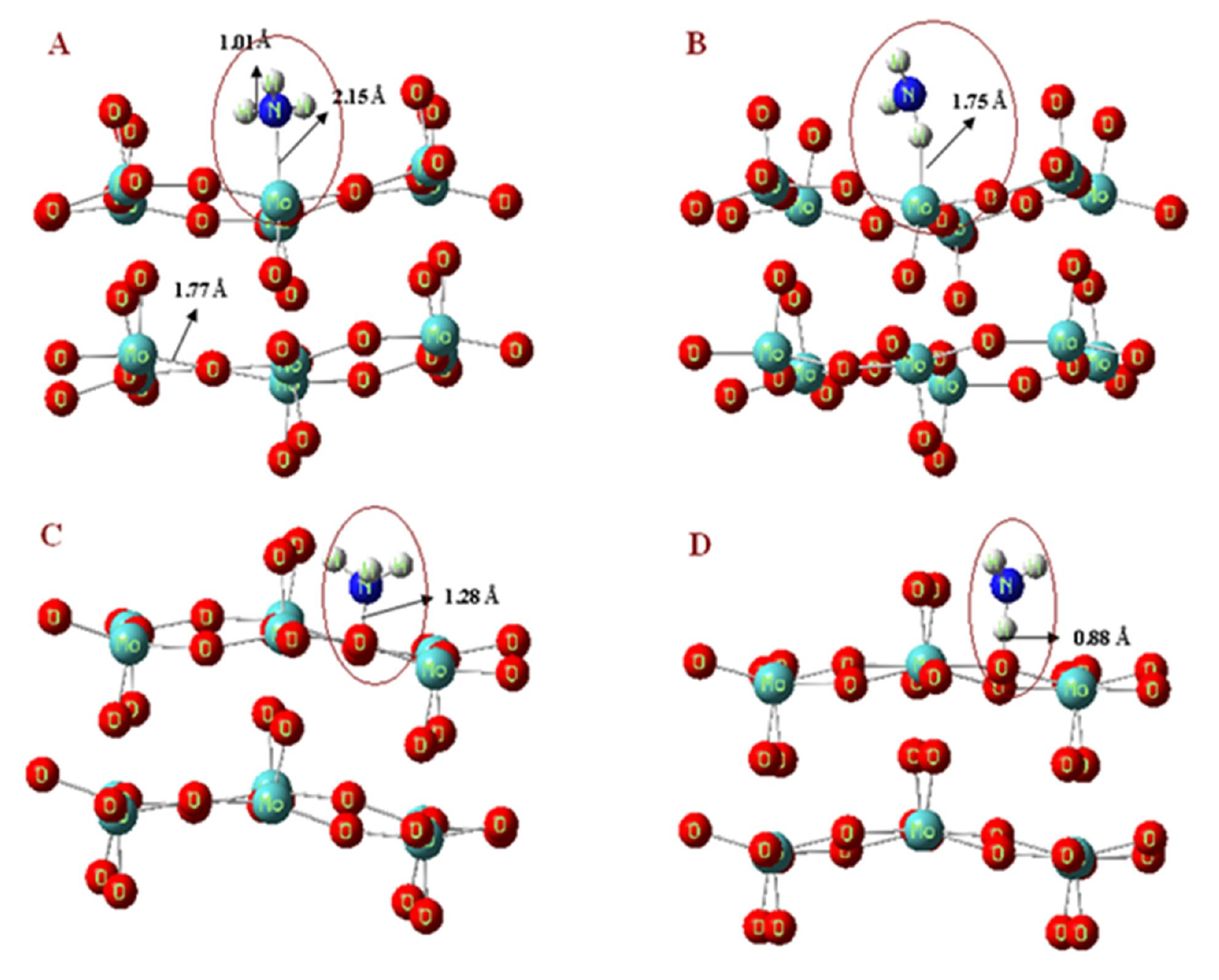}
\caption{(Color online) NH$_3$ adsorbed on position A, B, C and D in pristine MoO$_3$ nanostructure.} \label{fig7}
\end{center}
\end{figure}

\begin{figure}[!t]
\begin{center}
\includegraphics[width=0.8\textwidth]{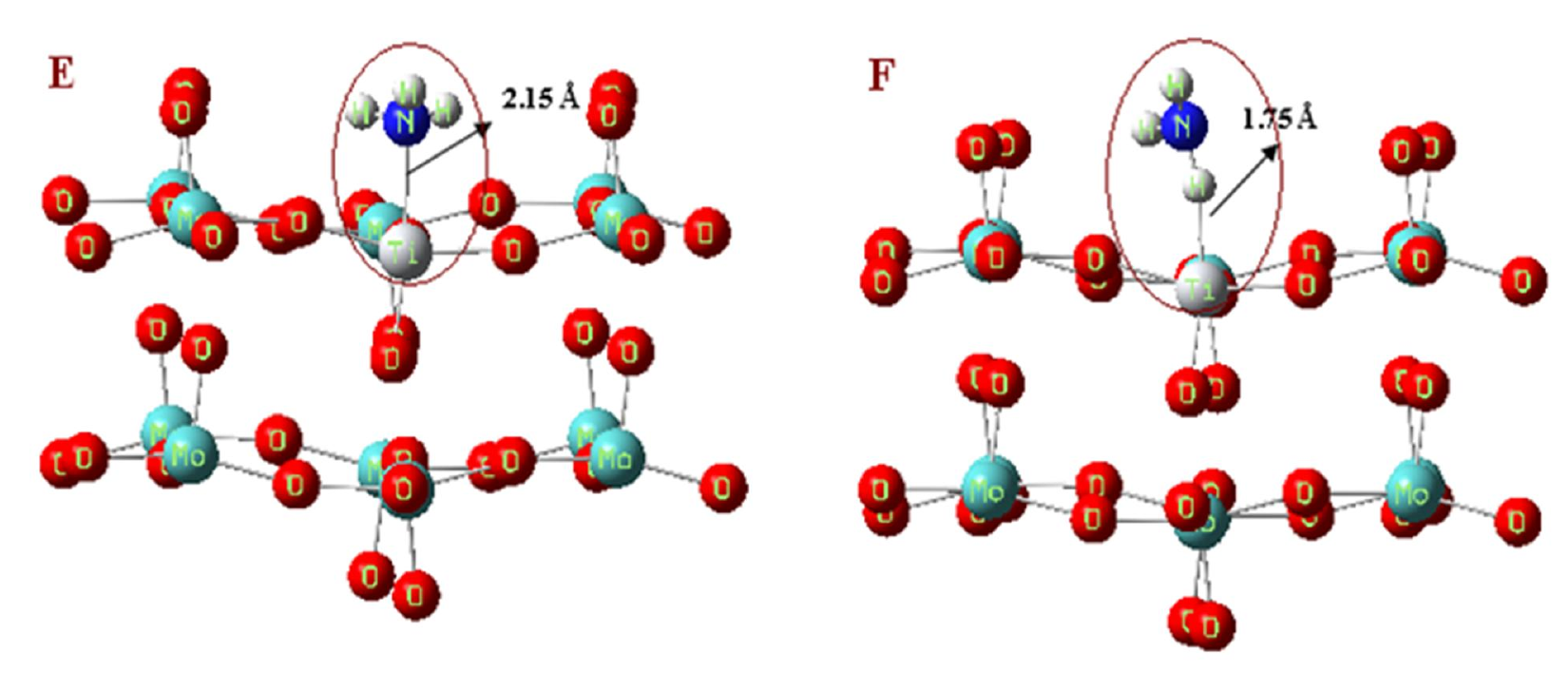}
\caption{(Color online) NH$_3$ adsorbed on position E and F in Ti substituted MoO$_3$ nanostructure.} \label{fig8}
\end{center}
\end{figure}

\begin{figure}[!t]
\begin{center}
\includegraphics[width=0.8\textwidth]{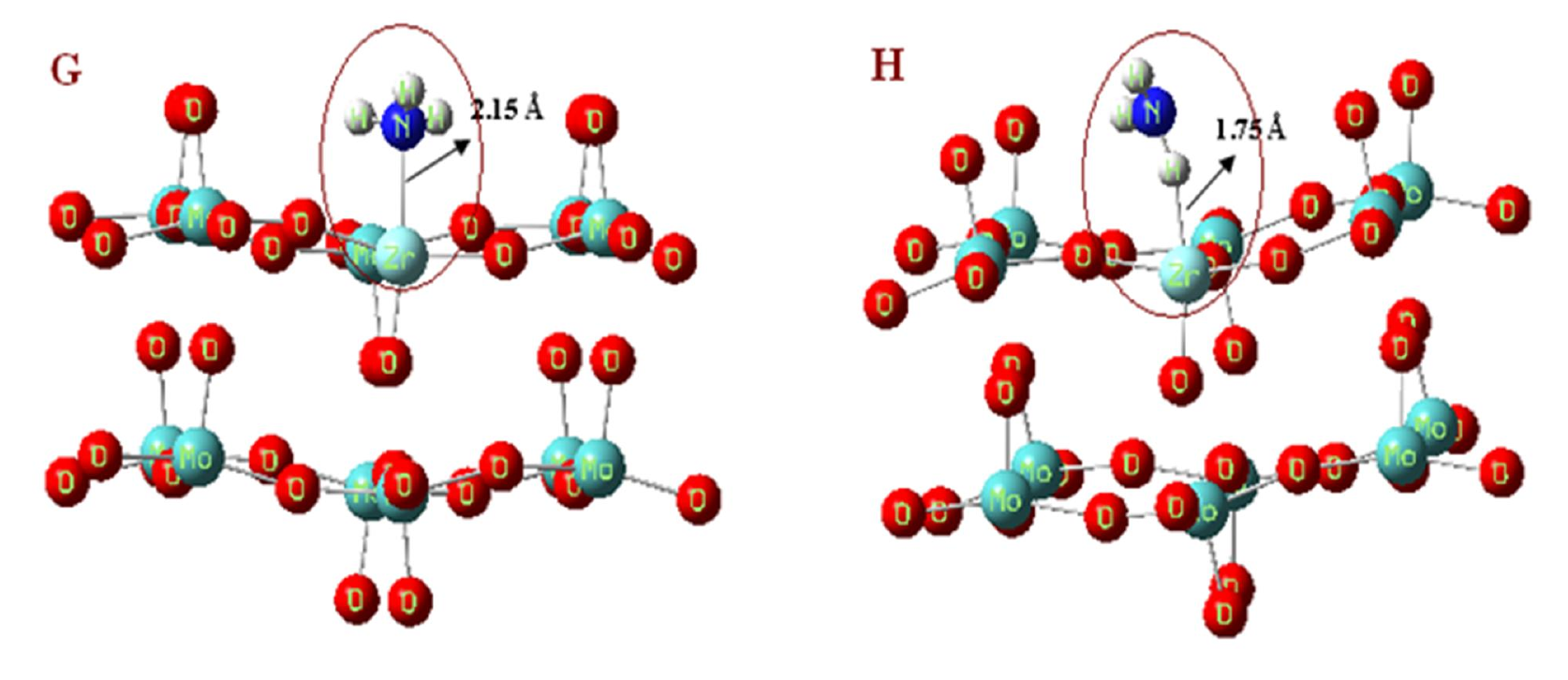}
\caption{(Color online) NH$_3$ adsorbed on position G and H in Zr substituted MoO$_3$ nanostructure.} \label{fig9}
\end{center}
\end{figure}

\begin{figure}[!t]
\begin{center}
\includegraphics[width=0.8\textwidth]{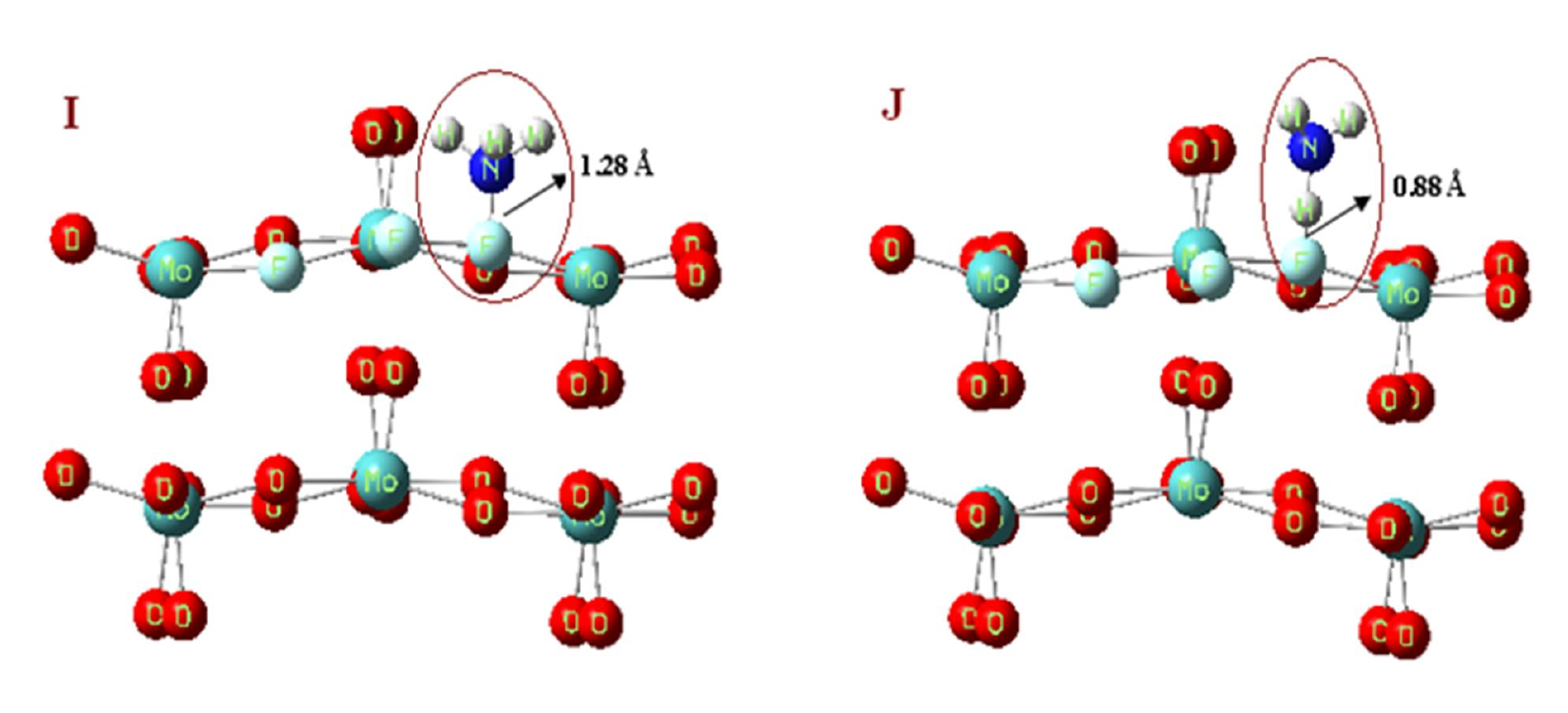}
\caption{(Color online) NH$_3$ adsorbed on position I and J in F substituted MoO$_3$ nanostructure.} \label{fig10}
\end{center}
\end{figure}

Before studying the adsorption properties of NH$_3$ on MoO$_3$ nanostructures, NH$_3$ should be investigated in gas phase. The optimized bond length between nitrogen and hydrogen atom in NH$_3$ is 1.01~{\AA} and the bond length between molybdenum and oxygen atoms in $\alpha$-MoO$_3$ is 1.77~{\AA}. These bond lengths are used during the optimization of $\alpha$-MoO$_3$ nanostructures. Figure~\ref{fig7} refers to the adsorption of  NH$_3$ gas molecule adsorbed on different sites, namely position A, B, C and D in pristine $\alpha$-MoO$_3$ nanostructure. Figures~\ref{fig8}--\ref{fig10} represent the adsorption of ammonia gas molecules adsorbed on various sites such as position E, F, G, H, I and J in Ti, Zr and F substituted $\alpha$-MoO$_3$ nanostructure.

The adsorption energy of NH$_3$ gas molecules on $\alpha$-MoO$_3$ nanostructure can be calculated by the equation~(\ref{3.2}) as follows:
\begin{equation}
E_{\text{ad}} = [E(\text{MoO}_3) + E(\text{NH}_3) - E(\text{MoO}_3/\text{NH}_3) + E(\text{BSSE})],	
\label{3.2}
\end{equation}
where $E$(MoO$_3$/NH$_3$) denotes the energy of MoO$_3$/NH$_3$ complex, $E$(MoO$_3$) and $E$(NH$_3$) are the isolated energies of MoO$_3$ and NH$_3$ molecules, respectively. The basis set superposition error (BSSE) \cite{50} can be analyzed in terms of counterpoise technique to eliminate the overlap effects on basis functions. $\alpha$-MoO$_3$ base material is found to have a negative value of energy which confirms the stability of the base material. Moreover, when NH$_3$ gas molecules are adsorbed on $\alpha$-MoO$_3$ nanostructures, negative values of the adsorption energy ($E_{\text{ad}}$) refer to the more stable system \cite{51,52}.

In the present work, positions B to J all have negative values of adsorption energy. This infers that $\alpha$-MoO$_3$ nanostructures are more stable and it is most suitable for gas sensor and as catalyst. The adsorption energies of pristine $\alpha$-MoO$_3$ for positions A, B, C and D are 1.09, $-$2.45, $-$8.16 and $-$11.7~eV, respectively. The corresponding $E_{\text{ad}}$ values of Ti incorporated $\alpha$-MoO$_3$ for positions E and F are $-$0.27 and $-$2.72~eV. Zr substituted $\alpha$-MoO$_3$ for positions G and H have $E_{\text{ad}}$ values of $-$1.09 and $-$3.26~eV, respectively. F substituted $\alpha$-MoO$_3$ for positions I and J has adsorption energy values of $-$14.42 and $-$9.52~eV, respectively. The other significant parameter to decide NH$_3$ adsorption characteristics is the band gap of $\alpha$-MoO$_3$ nanostructure. Conductivity of $\alpha$-MoO$_3$ nanostructure increases due to a decrease in the band gap, when NH$_3$ gets adsorbed on positions A, C, D, E, G and H of $\alpha$-MoO$_3$ nanostructures, the corresponding band gap values are 4.57, 3.2, 1.87, 1.35, 1.86 and 2.12~eV, which are lower than their corresponding isolated counterpart. By contrast, conductivity of $\alpha$-MoO$_3$ nanostructures decreases owing to an increase in the band gap, when NH$_3$ gets adsorbed on positions B, F, I and J with energy gap values of 4.71, 2.13, 2.71 and 3.47~eV, respectively, which are higher than their isolated counterpart.  From the observations, it is revealed that the adsorption energy and the energy gap values change due to the adsorption of NH$_3$ in $\alpha$-MoO$_3$ nanostructure.

Prasad et al. \cite{53} reported the gas-sensing properties of MoO$_3$ with ammonia gas. Imawan et al. \cite{19} proposed the gas-sensing properties of modified  MoO$_3$ thin films using Ti-overlayers for NH$_3$ gas sensors. With the influence of Ti-overlayers in MoO$_3$ material, selectivity and sensitivity of NH$_3$ gas molecules get enhanced. The response of NH$_3$ mainly depends upon temperature. Furthermore, to improve the adsorption of NH$_3$ on MoO$_3$, the activation energy is required. The high operating temperature of molybdenum oxide enhances the catalytic activity.  Therefore, the temperature dependent NH$_3$ gas sensitivity is possible in MoO$_3$ nanostructures. Based on the condition of the chemisorption mechanism, MoO$_3$ material shows good response to NH$_3$ gas molecules among other reducing gases such as H$_2$, CO and SO$_2$ at operating temperature of 200\degree{C}. Upon exposure of NH$_3$ gas molecules to molybdenum oxides, the resistance value changes leading to the detection of ammonia. The results obtained in the present work are for the ambient condition. Usually, by incorporating impurities or functionalization of the base material with Pd, Pt will decrease the operating temperature of metal oxide during gas sensing. Moreover, in the present study, the conductivity of $\alpha$-MoO$_3$ material also varies due to the transfer of electrons between NH$_3$ gas molecules and MoO$_3$ nanostructure. This validates the present work with the reported work.

In this work, the gas sensing properties of $\alpha$-MoO$_3$ nanostructures are studied with the substitution of Ti, Zr and F impurities. From the observation, it is revealed that the pristine, Ti, Zr and F incorporated $\alpha$-MoO$_3$ nanostructures are a promising material for sensing ammonia. The most favorable adsorption site of the NH$_3$ molecule on $\alpha$-MoO$_3$ material can be concluded only after investigating the variation in the average energy gap ($E_{\text g}^\text{a}$) along with its respective isolated counterpart \cite{54,55}. Table~\ref{table 2} refers the HOMO-LUMO gap, percentage variation in energy gap, adsorption energy and Mulliken population. As a result, it is inferred that the most suitable site for adsorption of NH$_3$ molecules on $\alpha$-MoO$_3$ nanostructures are positions C, D, E, G, I and J. The adsorption of N atom in NH$_3$ adsorbed on O, Ti, Zr and F atoms of pristine, Ti, Zr and F substituted $\alpha$-MoO$_3$ nanostructure and the adsorption of H atom in NH$_3$ adsorbed on O and F atom of pristine and F substituted $\alpha$-MoO$_3$ nanostructure are observed to be a more prominent adsorption site. The average energy gap variation is comparatively high among other adsorption sites.

\begin{figure}[!b]
\begin{center}
\includegraphics[width=0.81\textwidth]{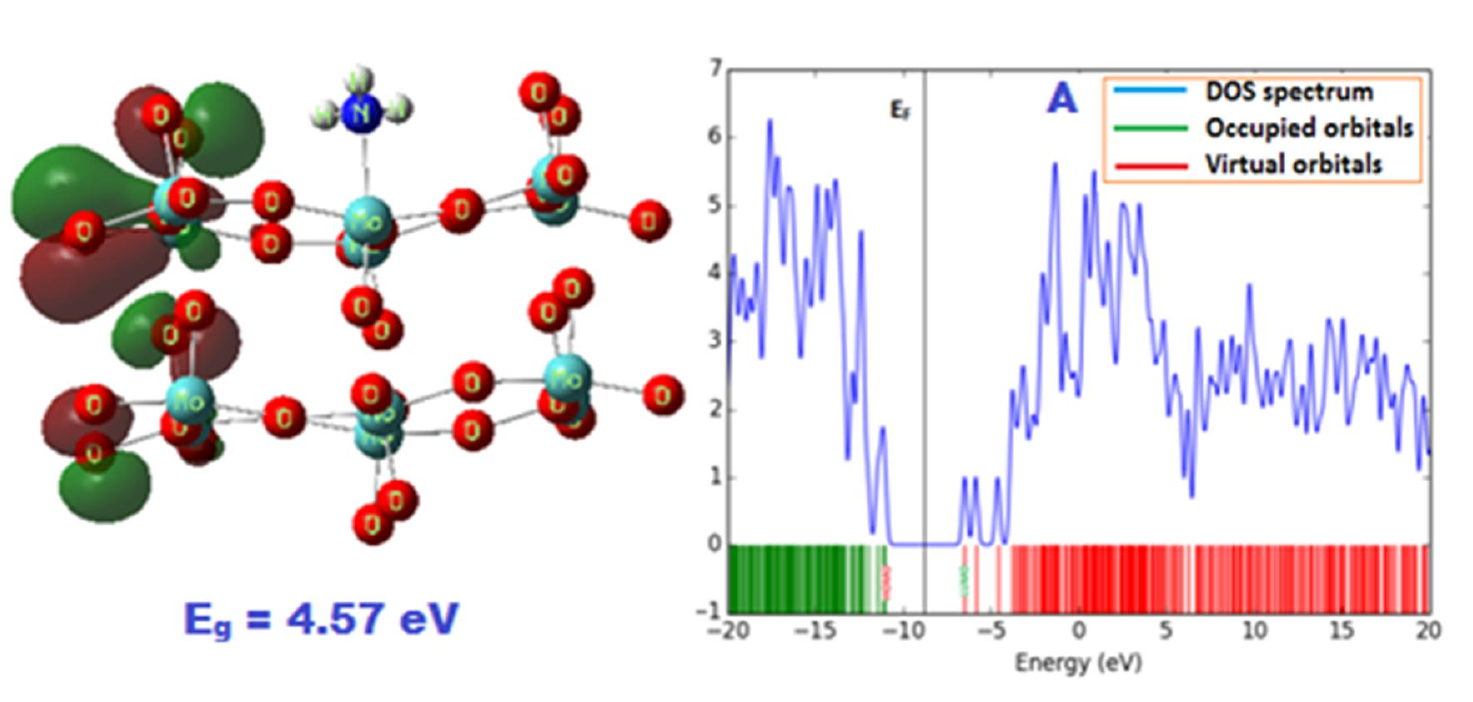}
\caption{(Color online) HOMO-LUMO gap and density of states of position A.} \label{fig11}
\end{center}
\end{figure}

\begin{figure}[!t]
\begin{center}
\includegraphics[width=0.81\textwidth]{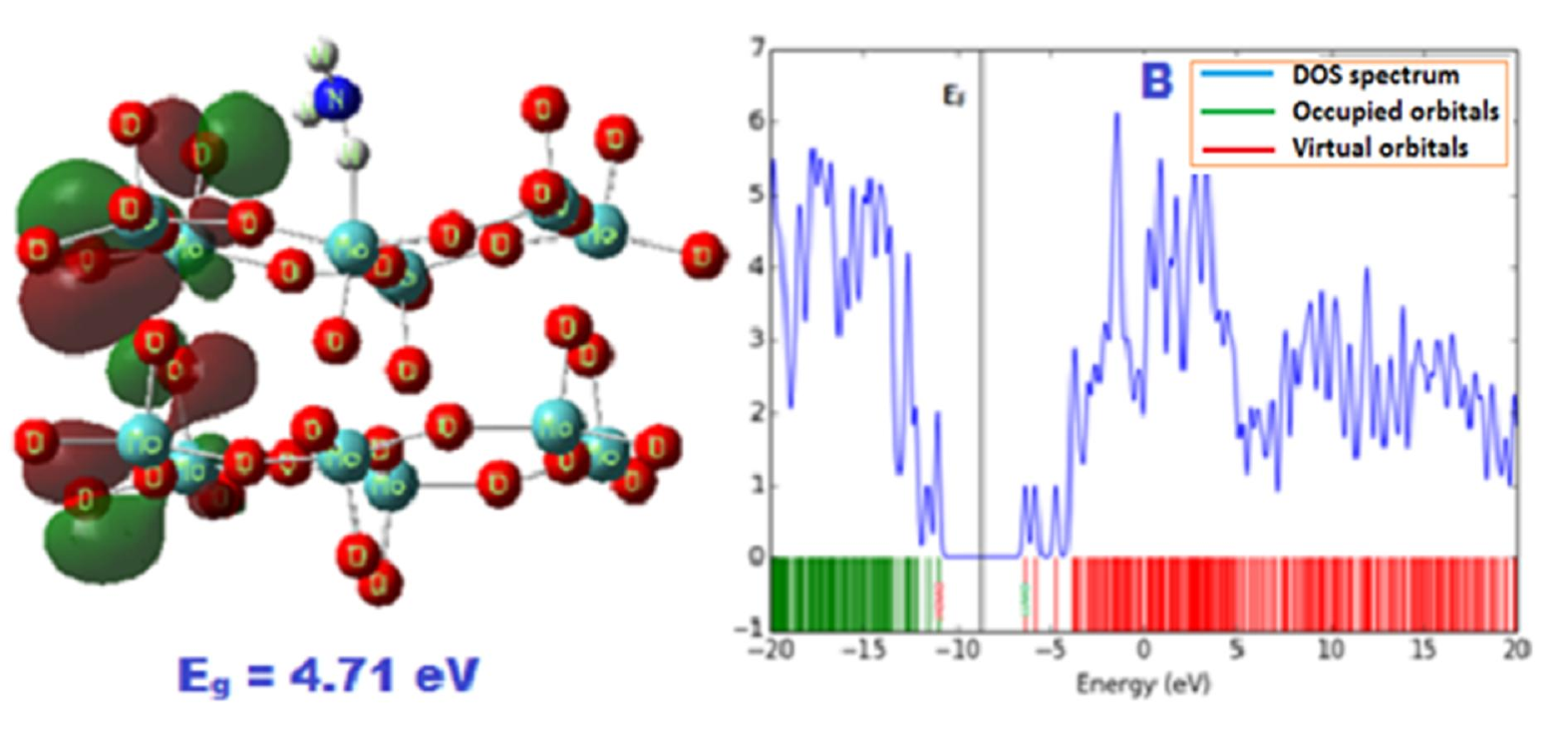}
\caption{(Color online) HOMO-LUMO gap and density of states of position B.} \label{fig12}
\end{center}
\end{figure}

\begin{figure}[!t]
\begin{center}
\includegraphics[width=0.81\textwidth]{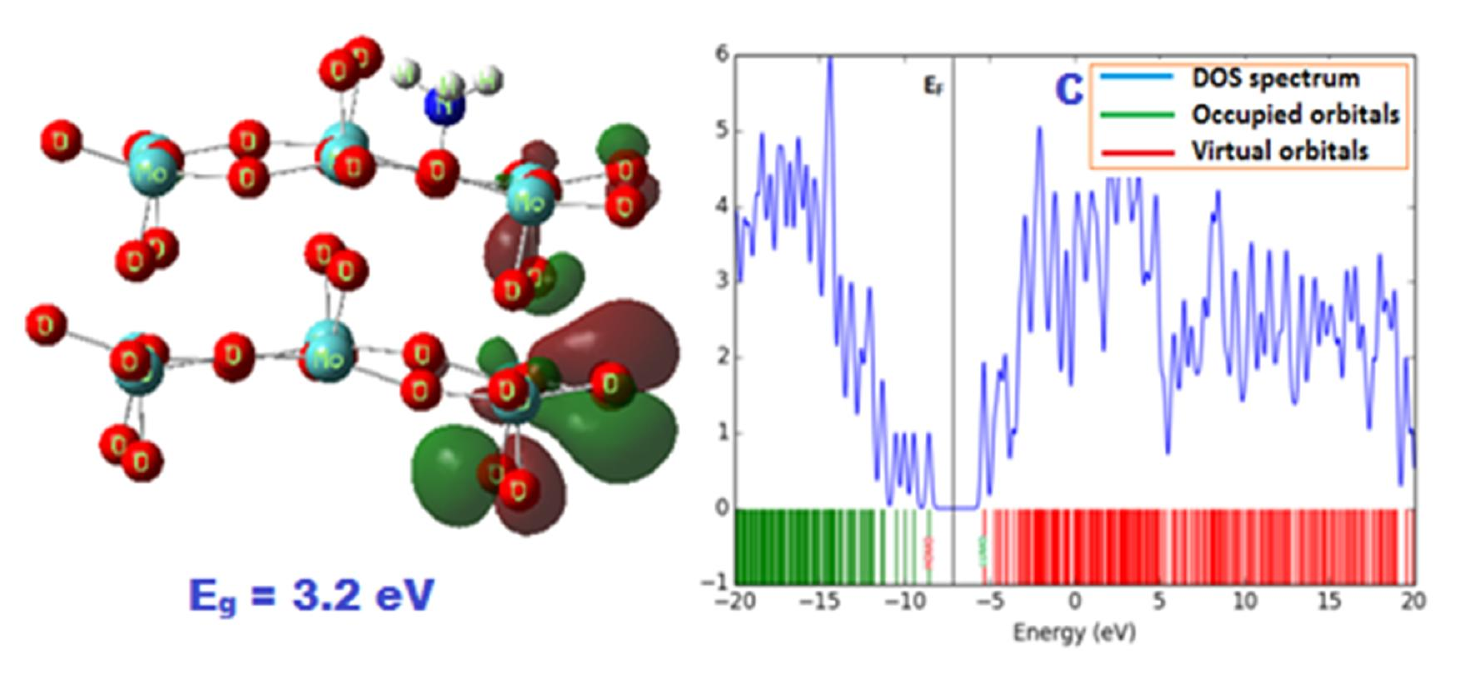}
\caption{(Color online) HOMO-LUMO gap and density of states of position C.} \label{fig13}
\end{center}
\end{figure}

\begin{figure}[!t]
\begin{center}
\includegraphics[width=0.81\textwidth]{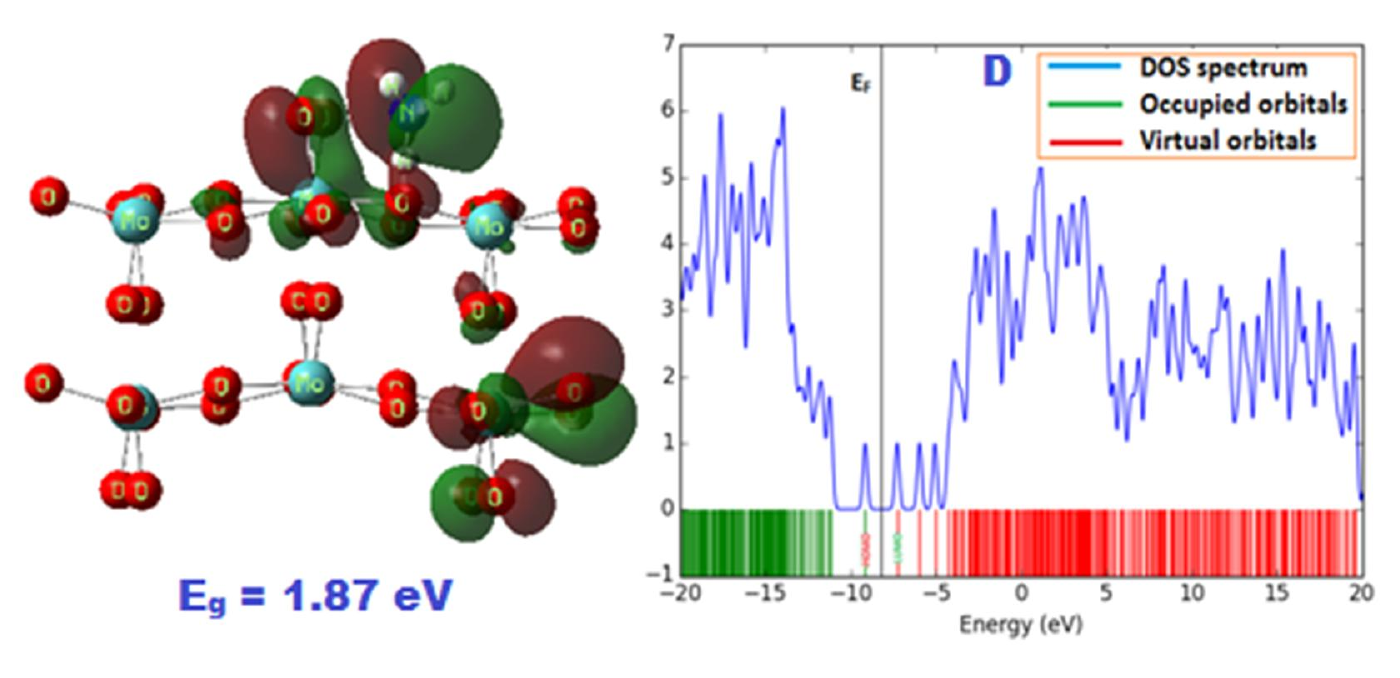}
\caption{(Color online) HOMO-LUMO gap and density of states of position D.} \label{fig14}
\end{center}
\end{figure}

\begin{figure}[!t]
\begin{center}
\includegraphics[width=0.81\textwidth]{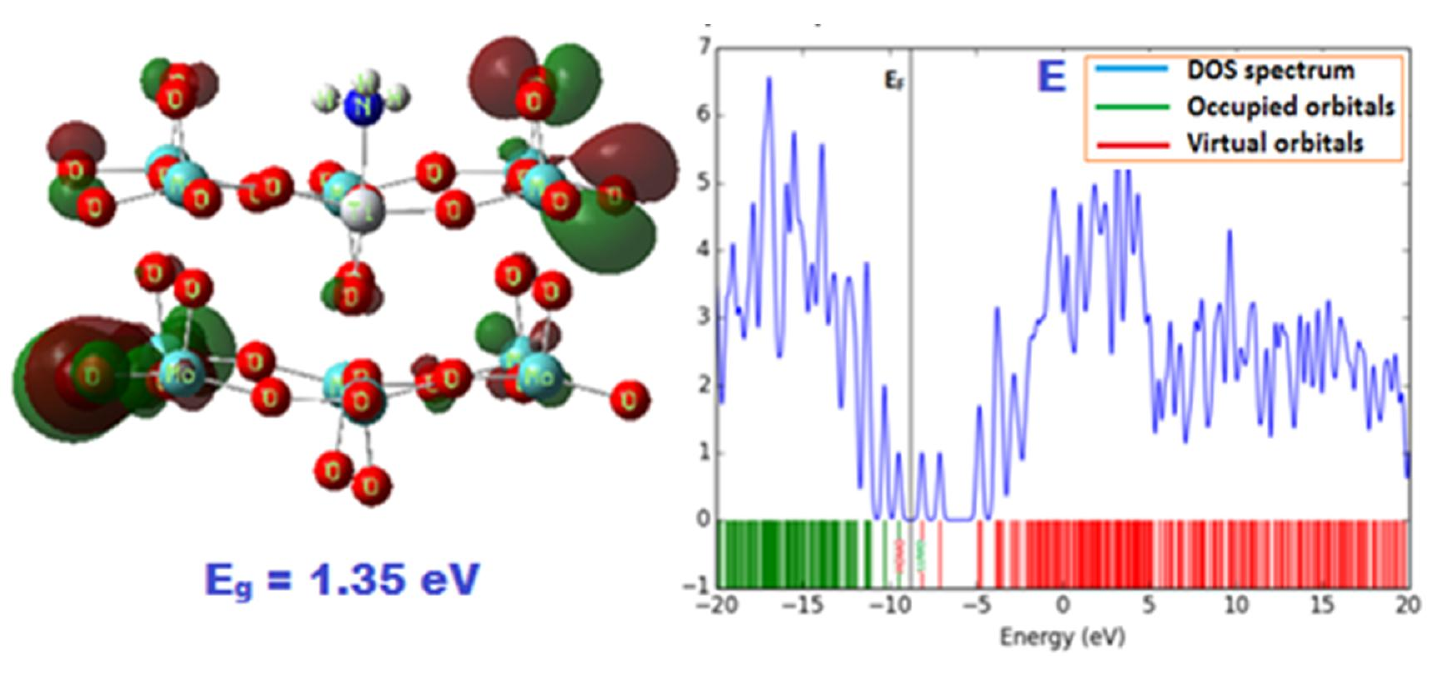}
\caption{(Color online) HOMO-LUMO gap and density of states of position E.} \label{fig15}
\end{center}
\end{figure}

\begin{figure}[!t]
\begin{center}
\includegraphics[width=0.81\textwidth]{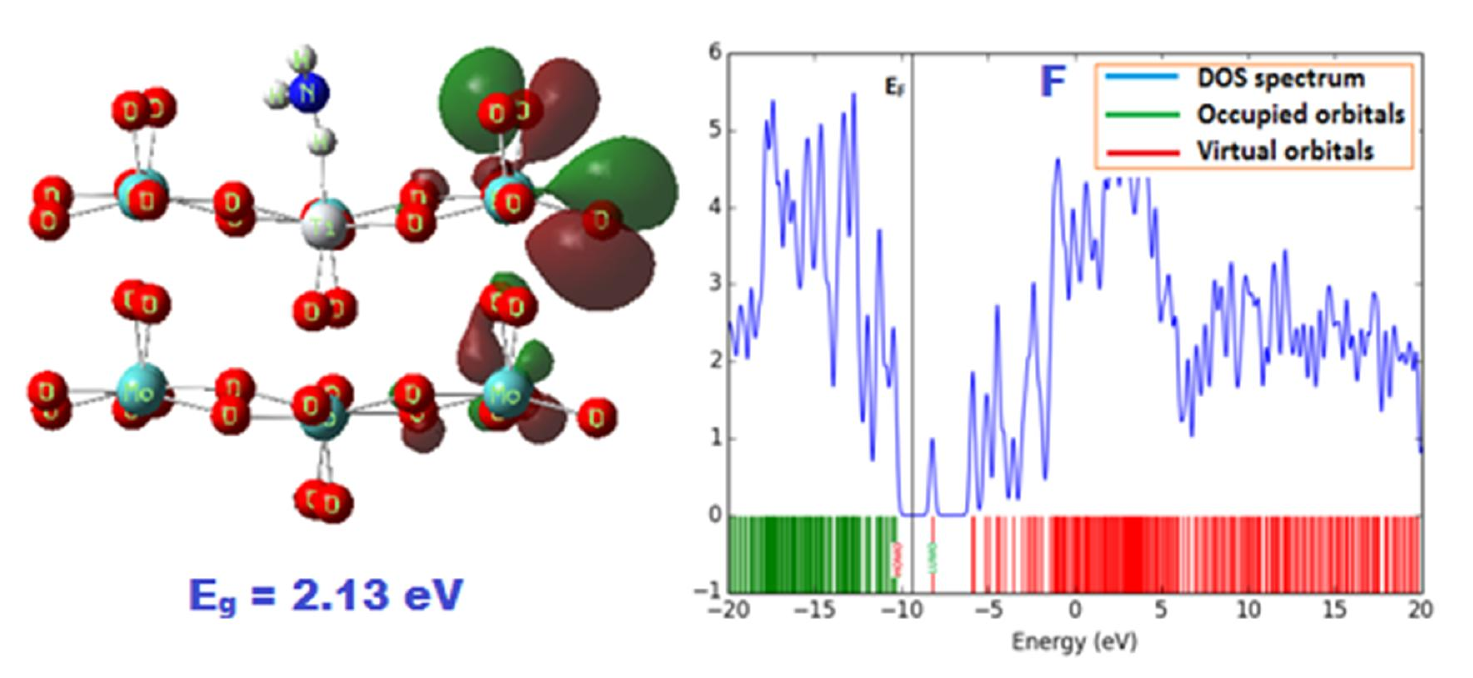}
\caption{(Color online) HOMO-LUMO gap and density of states of position F.} \label{fig16}
\end{center}
\end{figure}

\begin{figure}[!t]
\begin{center}
\includegraphics[width=0.81\textwidth]{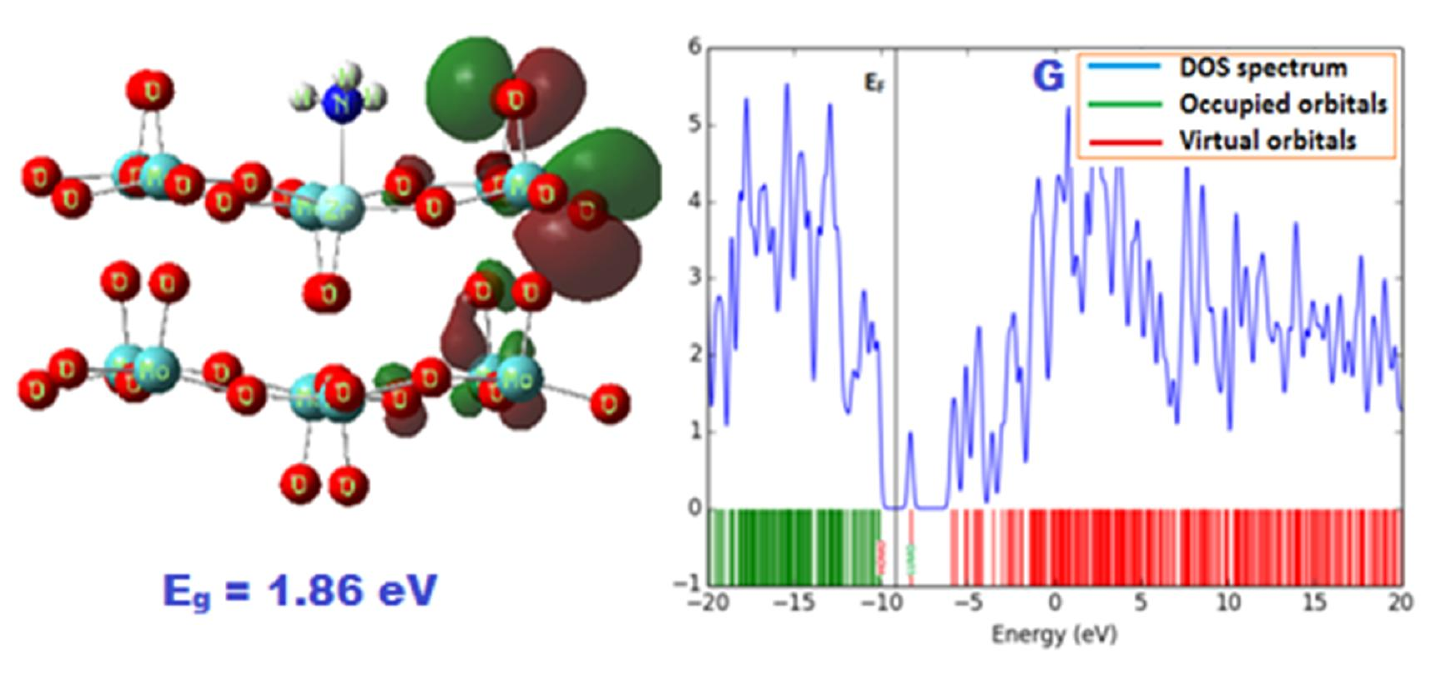}
\caption{(Color online) HOMO-LUMO gap and density of states of position G.} \label{fig17}
\end{center}
\end{figure}

\begin{figure}[!t]
\begin{center}
\includegraphics[width=0.81\textwidth]{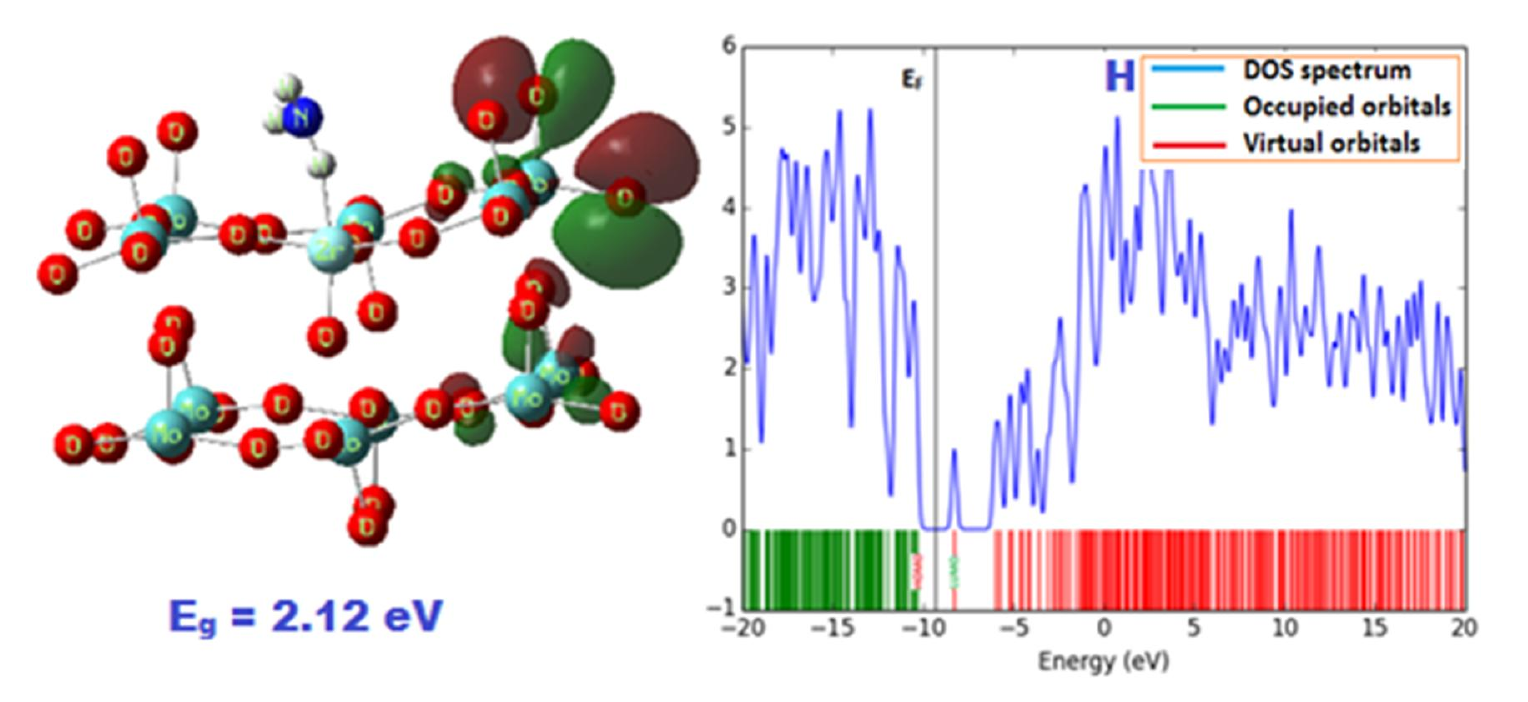}
\caption{(Color online) HOMO-LUMO gap and density of states of position H.} \label{fig18}
\end{center}
\end{figure}

\begin{figure}[!t]
\begin{center}
\includegraphics[width=0.81\textwidth]{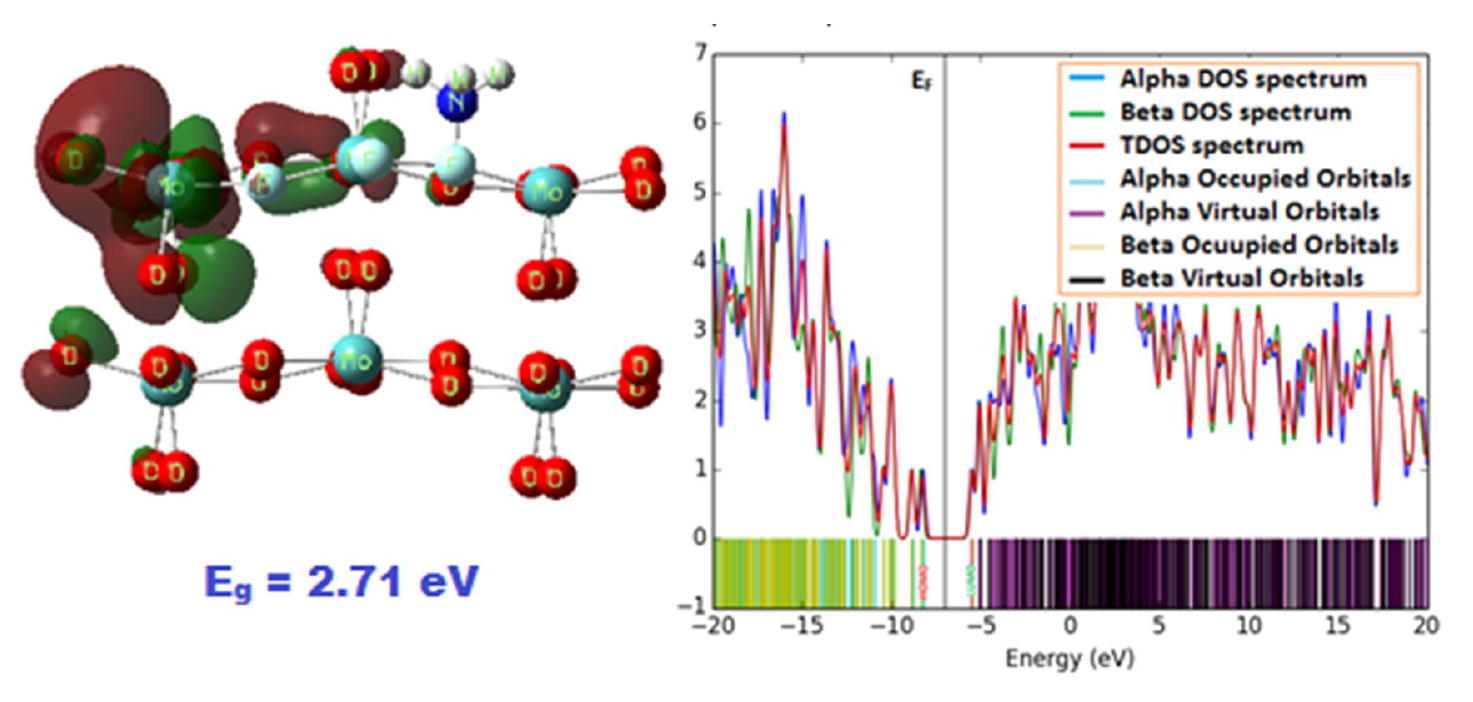}
\caption{(Color online) HOMO-LUMO gap and density of states of position I.} \label{fig19}
\end{center}
\end{figure}

\begin{figure}[!t]
\begin{center}
\includegraphics[width=0.81\textwidth]{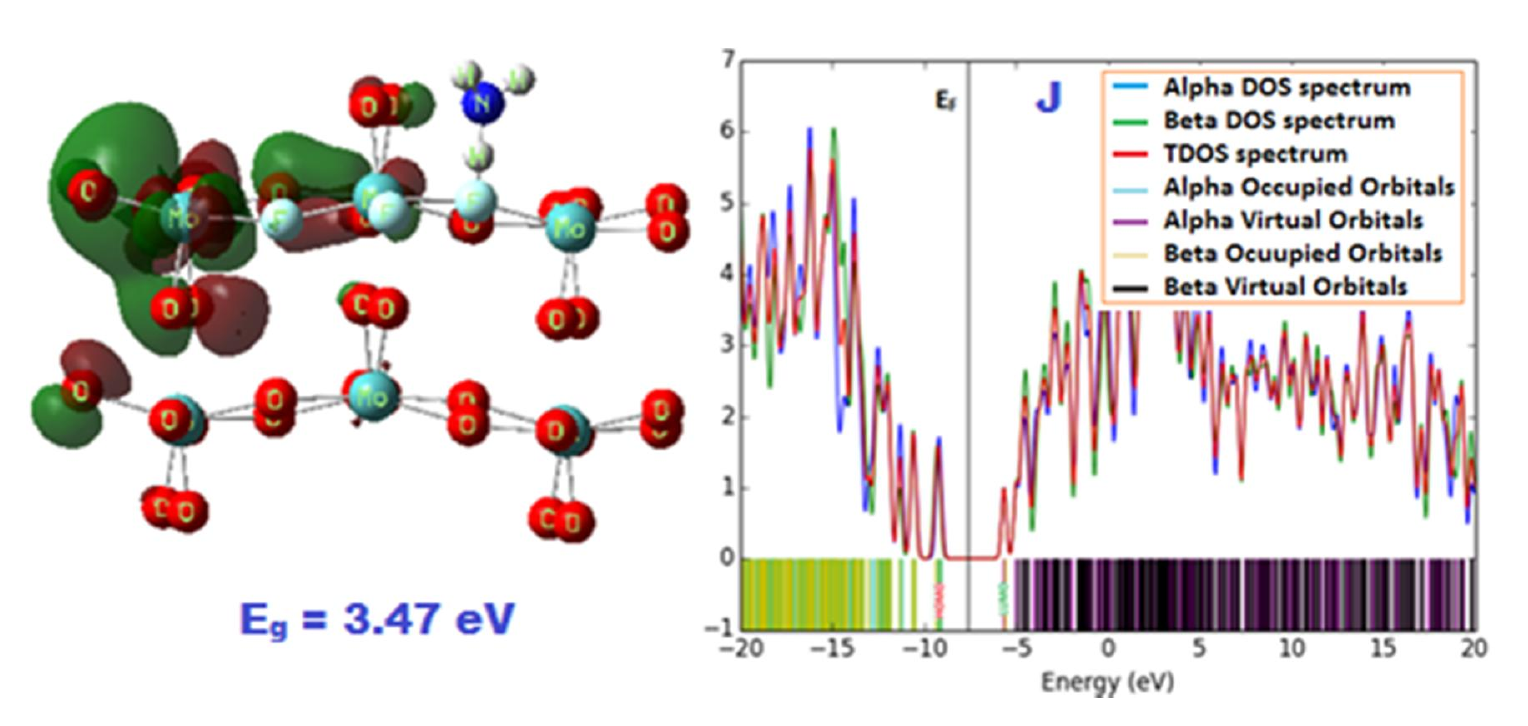}
\caption{(Color online) HOMO-LUMO gap and density of states of position J.} \label{fig20}
\end{center}
\end{figure}

The transfer of electrons between ammonia gas molecules and $\alpha$-MoO$_3$ can also be analyzed with the help of Mulliken population analysis ($Q$) \cite{56,57,58}. The negative charge of Mulliken population shows that the electrons are transferred from $\alpha$-MoO$_3$ material to NH$_3$ gas molecules while the positive value of $Q$ shows that the electrons are transferred from NH$_3$ gases to $\alpha$-MoO$_3$ base material. In the present work, all positions have positive value of $Q$, which infers that the electrons are transferred from NH$_3$ molecules to $\alpha$-MoO$_3$ nanostructure. The $Q$ values of pristine $\alpha$-MoO$_3$ nanostructure for positions A, B, C and D are 0.3, 0.11, 0.93 and 0.56~e, respectively. The corresponding Mulliken charge transfer values of Ti, Zr and F incorporated $\alpha$-MoO$_3$ nanostructure for positions E, F, G, H, I and J are 0.19, 0.10, 0.13, 0.08, 1.14 and 0.67~e. As a result, it is observed that the high value of Mulliken charge transfer, average energy gap variation and adsorption energy are found to be more prominent for positions C, D, E, G, I and J. Conductivity of $\alpha$-MoO$_3$ nanostructure for positions A, C, D, E, G and H increases owing to the narrowing of the energy gap. The remaining positions such as, B, F, I and J have less conductivity compared to the isolated counterpart. However, the conductivity decreases for positions I and J, the average energy gap variations are observed to be high, which is favorable for a good gas sensor.  Therefore, the most favorable adsorption sites can be found only after investigating the adsorption energy, Mulliken population and HOMO-LUMO gap of $\alpha$-MoO$_3$ nanostructure. Figures~\ref{fig11}--\ref{fig20} represent the density of states spectrum and HOMO-LUMO gap of position A--J of $\alpha$-MoO$_3$ nanostructures, respectively. Below the DOS spectrum, the green and red line indicates the HOMO-LUMO gap.  From the DOS spectrum, it is clearly revealed that more peak maxima are observed in unoccupied orbital than in occupied orbital.


This implies that the electrons can easily transfer between $\alpha$-MoO$_3$ and NH$_3$ gas molecules. Among all the optimum positions of $\alpha$-MoO$_3$ nanostructure, positions C, D, E and J have more peak maximum in unoccupied orbital.  Usually, in metal oxide based chemiresistors, which are used for gas/vapour sensing, the exchange of electrons takes place between the target gas and the base material. In the present study, a larger number of peak maxima is observed in the unoccupied orbital, which confirms the transfer of electrons between NH$_3$ target gas and $\alpha$-MoO$_3$ base material for different positions (C, D, E and J).  The alpha orbital and beta orbital exist for positions I and J, which arise due to spin up and down electrons due to orbital overlapping of F atoms with $\alpha$-MoO$_3$ nanostructure. This further strengthens NH$_3$ adsorption properties on $\alpha$-MoO$_3$ nanostructure, when substituted with F in $\alpha$-MoO$_3$ base material. Hence, the variation of the band gap leads to the change in the resistance of $\alpha$-MoO$_3$ nanostructure, which can be measured with a simple two probe arrangement. Analyzing all the aspects, it can be concluded that $\alpha$-MoO$_3$ can be used as NH$_3$ gas sensing material in the mixed gas atmosphere.

\section{Conclusions}
To sum up, DFT method is employed to study NH$_3$ adsorption properties on $\alpha$-MoO$_3$ material with B3LYP/LanL2DZ basis set. The electronic and structural stability of pristine, Ti, Zr and F substituted $\alpha$-MoO$_3$ nanostructure has been investigated. The structural stability of all $\alpha$-MoO$_3$ nanostructures has also been investigated using the formation energy. The electronic properties of $\alpha$-MoO$_3$ nanostructures are discussed in terms of HOMO-LUMO gap, ionization potential and electron affinity. The dipole moment and the point group of pristine, Ti, Zr and F substituted $\alpha$-MoO$_3$ nanostructures are also reported. The most prominent adsorption site of NH$_3$ on $\alpha$-MoO$_3$ nanostructure is identified and discussed in terms of adsorption energy, average energy gap variation, Mulliken population analysis and HOMO-LUMO gap. Furthermore, the negative values of adsorption energy for the positions B to J confirm the stability of $\alpha$-MoO$_3$ nanostructure upon adsorption of NH$_3$ molecules. The adsorption characteristics of NH$_3$ gas molecules get modified with the incorporation of Ti and F as impurities in $\alpha$-MoO$_3$ base material. Moreover, there is no significant improvement in NH$_3$ adsorption properties on $\alpha$-MoO$_3$ material with Zr substitution. In conclusion, the most favorable adsorption site of NH$_3$ on $\alpha$-MoO$_3$ nanostructures is when the N atom in NH$_3$ adsorbed on O, Ti, Zr and F atoms of pristine, Ti, Zr and F substituted $\alpha$-MoO$_3$ nanostructure and the adsorption of H atom in NH$_3$ adsorbed on the O and F atoms of pristine and F substituted $\alpha$-MoO$_3$ nanostructure. Thus, $\alpha$-MoO$_3$ nanostructure can be used as a good NH$_3$ gas sensor. In addition, the sensing characteristics of NH$_3$ can be modified with the incorporation of Ti and F as dopant in $\alpha$-MoO$_3$ nanostructures.

\ukrainianpart

\title{Взаємодія газу NH$_3$ на наноструктурах $\alpha$-MoO$_3$ --- дослідження за допомогою  теорії функціоналу густини}%
\author{В. Нагараджан, Р. Чандірамулі}
\address{Школа електротехніки та електроніки, Академія мистецтв, наукових і технологічних досліджень  Шанмуга (університет SASTRA), Танджавур, Таміл-Наду --- 613 401, Індія
}

\makeukrtitle

\begin{abstract}
Структурна стійкість, електронні властивості і  NH$_3$  адсорбційні властивості первинних,  Ti, Zr і F заміщених $\alpha$-MoO$_3$ наноструктур  успішно вивчені, використовуючи теорію функціоналу густини з B3LYP/ LanL2DZ базисним  набором. Структурна стійкість  $\alpha$-MoO$_3$ наноструктур
обговорюється в термінах енергії утворювання. Електронні властивості первинних,  Ti, Zr і F інкорпорованих  $\alpha$-MoO$_3$ наноструктур обговорюються в термінах HOMO-LUMO щілини, потенціалу іонізації та електронної афінності. $\alpha$-MoO$_3$ наноструктури
можуть бути точно-регульовані за допомогою підходящої домішки заміщення для покращення адсорбційних властивостей амоніяку, що може бути використано для виявлення  NH$_3$ в змішаному середовищі. Дана робота дає розуміння про застосування $\alpha$-MoO$_3$ наноструктур для виявлення  NH$_3$.
\keywords наноструктура, адсорбція, NH$_3$, HOMO-LUMO щілина, MoO$_3$ %
\end{abstract}

\end{document}